\documentclass[final,5p,times,twocolumn]{elsarticle}

\usepackage{amssymb}
\usepackage{amsmath}
\usepackage{hyperref}
\usepackage{booktabs}
\usepackage{array}
\usepackage{orcidlink}
\usepackage{tikz}
\usepackage{graphicx}
\usetikzlibrary{arrows.meta, positioning, fit, backgrounds, shapes.geometric, calc}

\journal{Kuwait Journal of Science}

\begin{document}

\begin{frontmatter}

\title{Boundary-Seeking GAN-Augmented TabTransformer for Adversarially Robust Intrusion Detection}

\author[ua]{Raihan Sultan Pasha Basuki\,\orcidlink{0009-0004-0010-4296}\corref{cor1}}
\ead{raihansultan.pashabasuki@students.uag.ac.id}

\author[ua]{Aliyah Kurniasih\, \orcidlink{0009-0002-7459-5230}}
\ead{aliyah.kurniasih@uag.ac.id}

\cortext[cor1]{Corresponding author.}
\affiliation[ua]{organization={Department of Computer Science, Universitas Ary Ginanjar},
            addressline={Menara 165, Cilandak No.kav 1 Lt 2 \& 18}, 
            city={Jakarta Selatan},
            postcode={12560}, 
            state={DKI Jakarta},
            country={Indonesia}}

\begin{abstract}
Machine learning-based intrusion detection systems (IDSs) often suffer from class imbalance and vulnerability to adversarial attacks, leading to degraded detection performance and reduced robustness. This study proposes a TabTransformer framework augmented by the Boundary-Seeking Generative Adversarial Network (BGAN) for flow-based intrusion detection using the CICIDS2017 dataset. BGAN serves a dual purpose by generating synthetic minority-class samples to mitigate data imbalance and producing adversarial samples to evaluate model robustness. Experimental results demonstrate that BGAN augmentation improves TabTransformer's Macro-F1 score from 82.96\% to 86.50\%, with the largest class-wise improvement observed for Web\_Attack (F1 score: 0.29 to 0.61). Robustness evaluation shows that all non-augmented models experienced a 100\% Performance Drop Rate (PDR) under adversarial testing, whereas all BGAN-augmented models achieved negative PDR values, indicating improved resilience. Furthermore, the augmented TabTransformer maintained stable and low FTR values (1.51\%--2.92\%) across all noise levels, compared with Decision Tree augmented with BGAN which reached 49.09\% under benign perturbations. These findings demonstrate that BGAN consistently enhances both class balance and adversarial robustness, while the proposed BGAN-TabTransformer framework provides an effective and adaptive intrusion detection solution for adversarial network environments.
\end{abstract}

\begin{highlights}
\item A BGAN-augmented TabTransformer is proposed for flow-based intrusion detection.
\item BGAN mitigates class imbalance and generates adversarial evaluation samples.
\item Macro-F1 improves from 82.96\% to 86.50\% on CICIDS2017.
\item BGAN augmentation substantially improves adversarial robustness.
\end{highlights}

\begin{keyword}
Intrusion Detection Systems \sep
Boundary-Seeking GAN \sep
TabTransformer \sep
Class Imbalance \sep
Adversarial Robustness

\end{keyword}

\end{frontmatter}

\section{Introduction}
\label{sec:intro}
 
The Internet has become the backbone of modern digital services, connecting education, finance, healthcare, and government infrastructure through open, high-speed data exchange. This openness, however, has made networks an attractive target for cyberattacks such as malware propagation, phishing, distributed denial-of-service (DDoS), data breaches, and network intrusion \cite{hnamte2024}, all of which can cause service disruption, sensitive data leakage, and substantial financial loss. Conventional network defenses -- firewalls, VPNs, and encryption -- remain largely reactive: firewalls block only traffic that violates predefined rules, and encryption cannot determine whether the traffic it protects is itself malicious \cite{alzaidy2024}. Intrusion Detection Systems (IDS) address this gap by passively monitoring and analyzing traffic patterns to flag anomalies without interrupting legitimate communication, making out-of-band deployment possible and enabling deeper post-hoc analysis of attack behavior.
 
While rule-based IDS are effective against known attack signatures, the constantly evolving threat landscape demands more adaptive detection. Machine learning (ML) and deep learning (DL) have consequently become the dominant paradigm for building IDS capable of recognizing previously unseen attack patterns \cite{hnamte2024}. This shift, however, introduces a new class of vulnerability: adversarial attacks. Alshahrani et al.~\cite{alshahrani2022} showed that GAN-based evasion and poisoning attacks can substantially degrade the accuracy of Decision Tree and Logistic Regression IDS models on the CICIDS2017 dataset, and more recent work has reported evasion rates against ML-based IDS as high as 99.9\% when adversarial examples are explicitly optimized to bypass detection. These findings expose a fundamental weakness: models that achieve high accuracy on clean data can collapse almost entirely when presented with subtly manipulated input, motivating IDS research that targets robustness alongside accuracy.
 
Prior work has advanced IDS design along two largely separate tracks. The first focuses on detection architecture. FlowLens \cite{barradas2021} classifies traffic using flow-marker distributions for fast anomaly detection, while pForest \cite{bussegrawitz2019} performs in-network inference with Random Forest for real-time classification; both, however, assume a stable traffic environment and do not account for deliberately manipulated adversarial traffic. More recently, Wang et al.~\cite{wang2024tabtransformer} applied a TabTransformer architecture to intrusion detection and reported an F1-score of 98.45\%, outperforming SVM, Logistic Regression, MLP, and voting ensembles by exploiting self-attention over tabular network features -- statistical flow attributes such as duration, packet size, byte counts, and TCP flags that carry meaning only in relation to one another rather than through spatial adjacency. That study, however, was limited to binary classification and did not evaluate performance under class imbalance or adversarial manipulation, both of which are characteristic of realistic network environments.
 
The second track uses Generative Adversarial Networks (GANs) either to synthesize minority-class samples for imbalance mitigation or to simulate adversarial traffic for robustness testing, but rarely both. IDSGAN \cite{lin2022idsgan} uses a Wasserstein-based generator to craft synthetic attack traffic that evades ML-based IDS; DRCGAN \cite{yue2025drcgan} adds residual connections to improve the stability of synthetic DDoS traffic under severe imbalance; and FR-GAN \cite{li2025frgan} incorporates a Soft Nearest Neighbor Loss to align synthetic and real feature distributions, improving minority-class F1-score without degrading majority-class performance. Complementary architectural advances, such as the multi-view heterogeneous graph model MH-Net~\cite{zhang2025mhnet} for encrypted traffic classification and deep neural network-based DDoS detection in software-defined networks~\cite{hnamte2024}, demonstrate the growing adoption of deep learning for network intrusion detection. Nevertheless, the increasing complexity of these models also enlarges the attack surface, as adversarial perturbations can significantly degrade detection performance~\cite{alzaidy2024}. Consequently, modern IDS should be both adaptive to evolving traffic patterns and resilient against adversarial manipulation.
 
Four limitations persist across this body of work. First, ML/DL-based IDS remain vulnerable to small-scale evasion perturbations of network features. Second, most GAN-assisted IDS research focuses primarily on offline model training rather than deployment in dynamic real-time detection environments. Third, class imbalance remains a fundamental obstacle: attack samples are vastly outnumbered by benign traffic, biasing models toward the majority class \cite{bagui2021}. Fourth, accurately modeling increasingly complex and encrypted network traffic remains a challenging task, motivating more expressive deep learning architectures such as MH-Net~\cite{zhang2025mhnet}. No existing framework combines a self-attention tabular classifier with a single generative model that serves \emph{both} as a minority-class augmentation engine and as an adversarial sample generator for robustness evaluation.
 
This study addresses that gap using a Boundary-Seeking GAN (BGAN), a GAN variant explicitly trained to generate samples near the decision boundary between benign and malicious classes \cite{ahmad2023bgan}. BGAN is applied for a dual purpose: (i) enriching minority-class samples to improve detection of rare attack types, and (ii) generating boundary-adjacent adversarial samples to stress-test model robustness -- a combination not addressed by augmentation-only frameworks such as FR-GAN or DRCGAN. For classification, we adopt TabTransformer \cite{huang2020tabtransformer}, whose self-attention mechanism is well suited to tabular flow features because it models feature-to-feature relationships directly. We evaluate on CICIDS2017 \cite{sharafaldin2018cicids}, a widely used IDS benchmark comprising roughly 2.8 million labeled flows across eight classes (Benign, Botnet, Brute Force, DDoS, DoS, Infiltration, Probe, and Web Attack) with severe class imbalance -- Infiltration accounts for only 36 samples, while Benign traffic makes up roughly 80\% of the dataset -- making it a suitable testbed for evaluating BGAN-based augmentation. Flow-statistical features were extracted using a flow-based representation consisting of packet counts, duration, inter-arrival times, packet-size statistics, and TCP flag information, consistent with flow-level IDS approaches such as FlowLens~\cite{barradas2021}.
 
The main contributions of this paper are as follows:
\begin{itemize}
\item We propose a dual-purpose BGAN-augmented TabTransformer framework for flow-based intrusion detection that simultaneously mitigates class imbalance and improves adversarial robustness, extending prior TabTransformer-based IDS work \cite{wang2024tabtransformer} from binary to multi-class classification under adversarial conditions.
\item We quantify the effect of BGAN augmentation on minority-class detection performance on CICIDS2017, with particular attention to the Botnet, Infiltration, and Web Attack classes.
\item We evaluate model robustness using Performance Drop Rate (PDR) under BGAN-generated adversarial samples and False Triggered Rate (FTR) under benign Gaussian perturbation, integrating augmentation and robustness evaluation within a single TabTransformer-based framework.
\item We show that BGAN augmentation improves Macro-F1 from 82.96\% to 86.50\% and converts a 100\% performance drop rate under adversarial testing into negative PDR values, indicating measurable robustness gains alongside improved class balance.
\end{itemize}
 
The overall framework is illustrated in Figure~\ref{fig:framework}. This study is scoped to model- and data-level evaluation in a controlled offline setting; in-network or on-switch deployment and evaluation on live real-time traffic are outside its scope.
 
The remainder of this paper is organized as follows. Section~\ref{sec:related} reviews related work on GAN-based augmentation and Transformer-based intrusion detection. Section~\ref{sec:methodology} describes the CICIDS2017 dataset, preprocessing pipeline, BGAN augmentation procedure, and TabTransformer architecture. Section~\ref{sec:results} presents classification and robustness results. Section~\ref{sec:discussion} discusses the implications and limitations of the findings, and Section~\ref{sec:conclusion} concludes the paper.
 
\section{Related Work}
\label{sec:related}
 
Prior research relevant to this study clusters around four themes: class-imbalance mitigation, adversarial robustness evaluation, GAN-based data augmentation for IDS, and Transformer-based intrusion detection architectures. This section reviews representative studies in each theme and positions the present work relative to them.
 
\textbf{Class-imbalance mitigation.} \cite{bagui2021} addressed the severe class imbalance present in benchmark IDS datasets such as KDD99 and UNSW-NB15 by applying resampling techniques -- random oversampling, random undersampling, their combination (RURO), SMOTE, and ADASYN -- prior to training an artificial neural network (ANN) classifier. Oversampling-based methods improved minority-class recall and Macro-F1 substantially (e.g., Macro-F1 on KDD99 fell from 0.843 without resampling to as low as 0.618--0.761 under several resampling variants, while recall rose from 0.833 to as high as 0.961), but at the cost of increased training time and a higher risk of overfitting. This illustrates a persistent trade-off in resampling-only approaches: minority-class sensitivity improves, but generalization and computational cost both suffer -- a trade-off that GAN-based synthetic augmentation aims to avoid by generating novel samples rather than duplicating or removing existing ones.
 
\textbf{Adversarial robustness evaluation.} \cite{alshahrani2022} were among the first to systematically evaluate ML-based IDS under GAN-generated adversarial conditions, testing Decision Tree (DT) and Logistic Regression (LR) classifiers on CICIDS2017 against both evasion and poisoning attacks crafted with a DCGAN. Evasion attacks reduced DT test accuracy, while poisoning attacks destabilized LR training; DT's F1-score fell from 60\% on clean data to 52\% under poisoning, showing that tree-based models are more susceptible to adversarial manipulation than linear models. This study established the CICIDS2017 adversarial-evaluation protocol adopted in the present work, but evaluated only binary classification with classical ML models, leaving open how a self-attention tabular architecture such as TabTransformer behaves under equivalent adversarial pressure.
 
\textbf{GAN-based augmentation for IDS.} Two studies motivate the specific choice of BGAN in this work. \cite{ahmad2023bgan} applied Boundary-Seeking GAN augmentation to classical ML models (Random Forest, Decision Tree, and ANN), reporting consistent gains: ANN F1-score rose from 79.78\% to 85.71\%, Decision Tree F1-score from 72.22\% to 84.86\%, and Random Forest precision from 72.43\% to 88.43\%. These results demonstrate that generating synthetic samples near the decision boundary -- rather than uniformly across the minority-class distribution -- is particularly effective for sharpening classifier discrimination, motivating BGAN's dual role in the present framework as both an augmentation and an adversarial-sample generator. \cite{li2025frgan} instead addressed a different failure mode of GAN augmentation: feature misalignment between synthetic and real samples. Their Feature-Regularized GAN (FR-GAN) adds a Soft Nearest Neighbor Loss to a WGAN backbone, improving Macro-F1 on CSE-CIC-IDS2018 from 75.9\% (SMOTE) to 84.6\% (WGAN-GP+SNNL) and on CICIDS2017 from 85.9\% (SMOTE) to 88.0\% (WGAN-GP+SNNL) while keeping accuracy unchanged. Neither study, however, uses the generator to construct adversarial test samples for robustness evaluation -- both treat augmentation and robustness as separate concerns, which is precisely the gap this paper addresses by using BGAN for both purposes within a single framework.
 
\textbf{Transformer-based intrusion detection.} \cite{wang2024tabtransformer} showed that TabTransformer's self-attention mechanism outperforms SVM, LR, MLP, and voting ensembles on tabular flow features, reaching an F1-score of 98.45\% versus 96.52\% for the next-best model (MLP), but the evaluation was restricted to binary classification on a military network traffic simulation without adversarial or imbalance testing. Related hybrid architectures illustrate the broader trend toward combining Transformers with GANs: \cite{salehiyan2025} proposed a Transformer--GAN--Autoencoder pipeline optimized with a Chimp Optimization Algorithm for edge and IIoT intrusion detection, achieving up to 98.92\% accuracy across three IIoT datasets (WUSTL-IIoT-2021, EdgeIIoTset, TON\_IoT), and \cite{chillara2025} combined a BERT Transformer with WGAN-GP augmentation to defend against USB keystroke-injection attacks, improving detection accuracy from 86.20\% to 90.02\%. Both confirm that Transformer--GAN combinations are effective, but neither targets flow-based network intrusion detection with an explicit, quantified robustness evaluation against adversarial perturbation -- the combination this paper provides.
 
Collectively, prior work shows that resampling and GAN-based
augmentation each improve minority-class detection in isolation,
that adversarial vulnerability has been demonstrated for classical
ML-based IDS \citep{alshahrani2022}, and that Transformer
architectures outperform prior classifiers on clean, balanced,
binary-classification tasks \citep{wang2024tabtransformer}.
No prior study, however, combines a self-attention tabular
classifier with a single dual-purpose generative model that
both augments minority classes \emph{and} generates adversarial
test samples for quantified robustness evaluation under
multi-class conditions -- the gap this paper addresses.
 
\section{Methodology}
\label{sec:methodology}

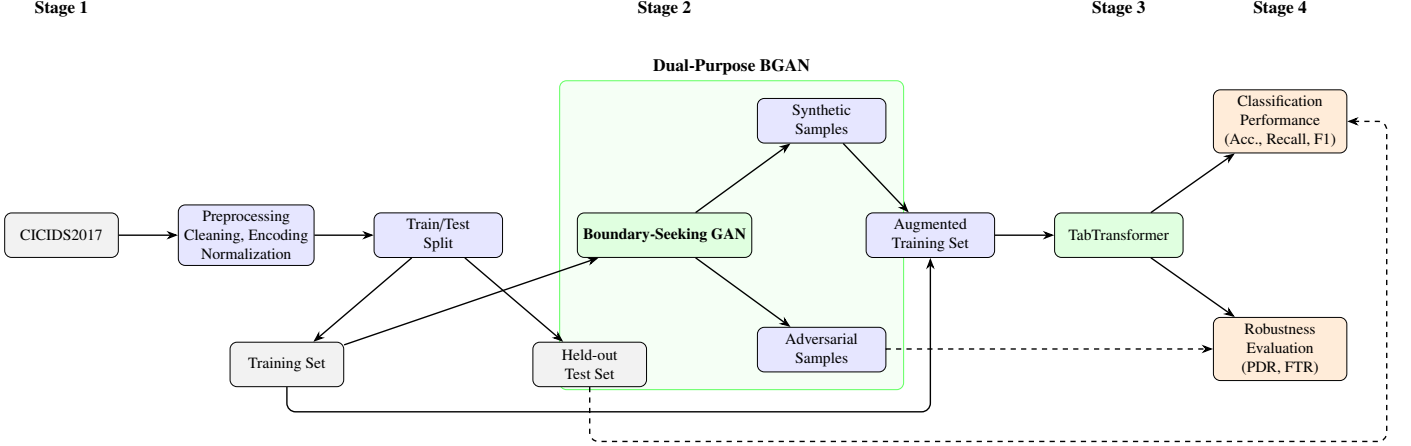
\begin{figure*}[t]
\centering
\resizebox{\textwidth}{!}{%
\begin{tikzpicture}[
font=\small,
>=Stealth,
dataset/.style={
    draw,
    rounded corners,
    fill=gray!10,
    minimum width=2.3cm,
    minimum height=0.9cm,
    align=center},
process/.style={
    draw,
    rounded corners,
    fill=blue!10,
    minimum width=2.6cm,
    minimum height=0.9cm,
    align=center},
model/.style={
    draw,
    rounded corners,
    fill=green!12,
    minimum width=2.6cm,
    minimum height=0.9cm,
    align=center},
eval/.style={
    draw,
    rounded corners,
    fill=orange!15,
    minimum width=2.7cm,
    minimum height=0.9cm,
    align=center},
arrow/.style={
    thick,
    ->}
]


\node[dataset] (dataset)
{CICIDS2017};

\node[process,right=1.2cm of dataset]
(pre)
{Preprocessing\\
Cleaning, Encoding\\
Normalization};

\node[process,right=1.2cm of pre]
(split)
{Train/Test\\Split};


\node[dataset,below left=1.7cm and 0.6cm of split]
(train)
{Training Set};

\node[dataset,below right=1.7cm and 0.6cm of split]
(test)
{Held-out\\Test Set};


\node[model,right=1.5cm of split]
(bgan)
{\textbf{Boundary-Seeking GAN}};

\node[process,above right=1.4cm and 0.1cm of bgan]
(aug)
{Synthetic\\Samples};

\node[process,below right=1.4cm and 0.1cm of bgan]
(adv)
{Adversarial\\Samples};


\node[process,right=2.3cm of bgan]
(merge)
{Augmented\\Training Set};

\node[model,right=1.2cm of merge]
(tab)
{TabTransformer};


\node[eval,above right=1.2cm and 0.6cm of tab]
(cls)
{Classification\\Performance\\(Acc., Recall, F1)};

\node[eval,below right=1.2cm and 0.6cm of tab]
(rob)
{Robustness\\Evaluation\\(PDR, FTR)};


\coordinate (route1) at ($(adv.south) + (0, -0.8)$);
\coordinate (route2) at ($(adv.south) + (0, -1.4)$);

\coordinate (TopLevel) at ($(aug.north) + (0, 1.8)$);


\draw[arrow] (dataset)--(pre);
\draw[arrow] (pre)--(split);

\draw[arrow] (split)--(train);
\draw[arrow] (split)--(test);

\draw[arrow] (train)--(bgan);

\draw[arrow] (bgan)--(aug);
\draw[arrow] (bgan)--(adv);

\draw[arrow] (aug)--(merge);
\draw[arrow] (merge)--(tab);
\draw[arrow] (tab)--(cls);
\draw[arrow] (tab)--(rob);

\draw[arrow, rounded corners=5pt]
(train.south) -- (train.south |- route1) -| (merge.south);

\draw[dashed,->, thick, rounded corners=5pt]
(test.south) -- (test.south |- route2) -| ($(cls.east) + (0.8,0)$) -- (cls.east);

\draw[dashed,->, thick]
(adv.east) |- (rob.west);


\begin{scope}[on background layer]

\node[
    draw=green!60,
    rounded corners,
    line width=0.6pt,
    fill=green!4,
    fit=(bgan)(aug)(adv),
    inner sep=10pt,
    label={[font=\bfseries]above:Dual-Purpose BGAN}
]{};

\end{scope}


\node[font=\bfseries] at (dataset |- TopLevel) {Stage 1};
\node[font=\bfseries] at (bgan |- TopLevel) {Stage 2};
\node[font=\bfseries] at (tab |- TopLevel) {Stage 3};
\node[font=\bfseries] at (cls |- TopLevel) {Stage 4};

\end{tikzpicture}
}
\caption{
Overview of the proposed BGAN-augmented TabTransformer framework.
After preprocessing and dataset partitioning, the training set is used to train a Boundary-Seeking GAN (BGAN). The trained BGAN serves a dual purpose by generating synthetic minority-class samples for data augmentation and adversarial samples for robustness evaluation. The augmented dataset is then used to train the TabTransformer classifier, while the held-out test set and adversarial samples are employed to evaluate classification performance and adversarial robustness, respectively.
}
\label{fig:framework}
\end{figure*}

\subsection{Dataset and Preprocessing}
\label{subsec:dataset}
The proposed framework is shown in Figure~\ref{fig:framework}.
This study uses CICIDS2017 \cite{sharafaldin2018cicids}, a network traffic dataset capturing five days of realistic traffic containing benign flows alongside DoS, DDoS, brute force, web attack, botnet, and infiltration activity, each flow described by 78 statistical features. The original 15 fine grained labels were regrouped into eight classes (Benign, DoS, DDoS, Probe, Brute\_Force, Botnet, Web\_Attack, Infiltration), and Benign traffic was undersampled to 200,000 samples to reduce majority class dominance, following \cite{salehiyan2025}.

The dataset was split into training and test sets using a stratified 80:20 ratio, so that minority classes such as Infiltration (29 training, 7 test samples) and Botnet (1,565 training, 391 test samples) remain proportionally represented. The test set was held out before augmentation and was not used to train any model or generator. Features were scaled with a StandardScaler fitted only on the training partition. Note that Benign undersampling was applied to the training partition only; the test set retains the original class proportions, yielding 454,265 Benign test samples.

Table~\ref{tab:distribution} summarizes the class distribution
of the training set before and after augmentation under each
scenario. Only the three targeted minority classes receive
synthetic samples; all other classes remain unchanged.

\begin{table}[t]
\centering
\caption{Training set class distribution before and after
augmentation. $\Delta$ Full = change under Full BGAN.}
\label{tab:distribution}
\scriptsize
\renewcommand{\arraystretch}{1.15}
\setlength{\tabcolsep}{3pt}
\begin{tabular}{lrrrr}
\toprule
\textbf{Class} & \textbf{Original} &
\textbf{Full BGAN} & \textbf{Mixed} &
\textbf{$\Delta$ Full} \\
\midrule
Benign       & 200,000 & 200,000 & 200,000 & 0       \\
Botnet       &   1,565 &   6,000 &   6,000 & +4,435  \\
Brute\_Force  &  11,066 &  11,066 &  11,066 & 0       \\
DDoS         & 102,420 & 102,420 & 102,420 & 0       \\
DoS          & 201,378 & 201,378 & 201,378 & 0       \\
Infiltration &      29 &   6,000 &     500 & +5,971  \\
Probe        & 127,043 & 127,043 & 127,043 & 0       \\
Web\_Attack   &   1,744 &   6,000 &   6,000 & +4,256  \\
\midrule
\textbf{Total} & \textbf{645,245} & \textbf{659,907}
  & \textbf{654,407} & \textbf{+14,662} \\
\bottomrule
\end{tabular}
\end{table}

The severity of class imbalance is evident: Infiltration
comprises only 29 training samples against 201,378 DoS
samples, a ratio exceeding 6,900:1. Botnet and Web\_Attack,
with 1,565 and 1,744 samples respectively, are less extreme
but still substantially underrepresented relative to Benign
traffic. Full BGAN augmentation increases the total training
set by 14,662 samples (2.27\%), a modest absolute increase
that nonetheless has a disproportionate effect on minority-class
detection, as shown in Table~\ref{tab:minority-comparison}.

\subsection{BGAN-Based Data Augmentation}
\label{subsec:bgan}

A standard GAN trains a generator $G$ and discriminator $D$ adversarially, but this objective is prone to vanishing gradients and is poorly suited to tabular data, where small perturbations near class boundaries matter more than overall realism. Boundary-Seeking GAN (BGAN) \cite{ahmad2023bgan} addresses this by steering the generator toward samples that lie near the discriminator's decision boundary rather than toward the bulk of the class distribution, using an importance weight derived from the discriminator output:

\begin{equation}
w(x) = \frac{D(x)}{1 - D(x)}
\label{eq:importance-weight}
\end{equation}

Samples the discriminator finds ambiguous ($D(x) \approx 0.5$) are emphasized during training, which improves training stability and makes BGAN suitable both for augmenting rare attack classes and for generating adversarial samples that probe the decision boundary directly, its dual role in this study.

The generator and discriminator are simple feedforward networks (two hidden layers of 128 units, trained with Adam for 200 epochs). BGAN was trained separately on the three minority classes, Botnet, Web\_Attack, and Infiltration. Because Infiltration has only 29 training samples, the generator could not learn its distribution (Wasserstein distance of 8.62 between real and synthetic samples), so a Gaussian noise fallback ($\sigma=0.05$) was used for this class instead. Three scenarios were evaluated: \textit{Baseline} (no augmentation), \textit{Full BGAN} (all three minority classes augmented with BGAN to 6,000 synthetic samples each), and \textit{Mixed} (BGAN for Botnet and Web\_Attack, Gaussian fallback for Infiltration). Synthetic samples were used only in training and never entered the test set.

Table~\ref{tab:synth-quality} reports the Wasserstein distance
between real and synthetic feature distributions for each class
and scenario, used to assess the quality of generated samples.

\begin{table}[t]
\centering
\caption{Synthetic data quality by class and augmentation mode.
Wasserstein $<$ 1.5: OK; $\geq$ 3.0: BAD.}
\label{tab:synth-quality}
\scriptsize
\renewcommand{\arraystretch}{1.15}
\setlength{\tabcolsep}{4pt}
\begin{tabular}{llccc}
\toprule
\textbf{Class} & \textbf{Mode} & \textbf{Method} &
\textbf{Wasserstein} & \textbf{Status} \\
\midrule
Botnet      & Full BGAN & BGAN     & 0.921 & OK  \\
Botnet      & Mixed     & BGAN     & 0.923 & OK  \\
Web\_Attack  & Full BGAN & BGAN    & 0.905 & OK  \\
Web\_Attack  & Mixed     & BGAN    & 0.574 & OK  \\
Infiltration & Full BGAN & BGAN   & 8.619 & BAD \\
Infiltration & Mixed    & Gaussian & 0.033 & OK  \\
\bottomrule
\end{tabular}
\end{table}

Botnet and Web\_Attack generators converge well under both
scenarios (Wasserstein $<$ 1.0), confirming that BGAN
successfully models the boundary region for these classes.
Infiltration's BGAN generator fails to produce a useful
distribution (Wasserstein $=$ 8.619), directly confirming
that 29 training samples are insufficient for stable GAN
convergence. The Gaussian fallback in the Mixed scenario
reduces this to 0.033, validating its use as a replacement
for this class specifically.

\subsection{TabTransformer Classifier}
\label{subsec:tabtransformer}

Classification uses TabTransformer \cite{huang2020tabtransformer}, which applies self-attention across tabular features so that each feature is represented in context of every other feature, rather than independently as in a standard feedforward network. Since all 78 CICIDS2017 features are continuous, each is projected directly into a 32-dimensional embedding, following the adaptation used by \cite{wang2024tabtransformer}. These embeddings pass through two Transformer blocks of multi-head self-attention,

\begin{equation}
\text{Attention}(Q,K,V) = \text{softmax}\left(\frac{QK^T}{\sqrt{d_k}}\right)V
\label{eq:attention}
\end{equation}

followed by a feedforward sublayer, before a multilayer perceptron head produces the final classification. The model was trained with AdamW (learning rate 0.001, batch size 256) and early stopping on a held-out validation split, separately for each of the three augmentation scenarios.

\subsection{Baseline Models}
\label{subsec:baselines}

Four baselines were trained under the same Baseline and Full BGAN scenarios for comparison: Random Forest and Decision Tree (both with balanced class weighting, matching the comparison models used in \cite{li2025frgan} and \cite{ahmad2023bgan}), a shallow ANN following \cite{ahmad2023bgan}'s configuration, and a deeper MLP that, unlike TabTransformer, processes features independently, serving as an ablation control for the effect of self-attention.

\subsection{Evaluation Metrics}
\label{subsec:robustness}

Classification performance is evaluated using Accuracy, Precision, Recall, per-class F1-score, and Macro-F1 on the held-out test set. These metrics are computed from the confusion matrix, where True Positive (TP), True Negative (TN), False Positive (FP), and False Negative (FN) denote the numbers of correctly and incorrectly classified samples.

\begin{equation}
\text{Accuracy} =
\frac{TP+TN}{TP+TN+FP+FN}
\label{eq:accuracy}
\end{equation}

\begin{equation}
\text{Precision} =
\frac{TP}{TP+FP}
\label{eq:precision}
\end{equation}

\begin{equation}
\text{Recall} =
\frac{TP}{TP+FN}
\label{eq:recall}
\end{equation}

\begin{equation}
F_1 =
2 \cdot
\frac{\text{Precision}\times\text{Recall}}
{\text{Precision}+\text{Recall}}
\label{eq:f1}
\end{equation}

For multi-class classification, the overall performance is summarized using the Macro-F1 score, which is defined as the arithmetic mean of the class-wise F1-scores:

\begin{equation}
\text{Macro-F1}
=
\frac{1}{C}
\sum_{i=1}^{C}F_{1,i}
\label{eq:macrof1}
\end{equation}

where $C$ denotes the total number of classes.

In addition to conventional classification metrics, adversarial robustness is evaluated using two complementary measures. Performance Drop Rate (PDR) quantifies the relative degradation in F1-score when clean test samples are replaced with BGAN-generated adversarial samples of the same class:

\begin{equation}
\text{PDR} = \frac{F1_{\text{clean}} - F1_{\text{adversarial}}}{F1_{\text{clean}}} \times 100\%
\label{eq:pdr}
\end{equation}

A positive PDR means performance degrades under attack; a negative PDR means it holds or improves. False Triggered Rate (FTR) measures how often correctly classified Benign samples are misclassified after small Gaussian perturbations ($\sigma \in \{0.01, 0.05, 0.10\}$) are added, i.e. the false alarm rate under harmless noise:

\begin{equation}
\text{FTR} = \frac{FP_{\text{perturbed}}}{N_{\text{Benign}}} \times 100\%
\label{eq:ftr}
\end{equation}

All models were implemented in Python with PyTorch and scikit-learn, and trained on a GPU-equipped server.

\section{Results}
\label{sec:results}

All results below are computed on the held-out test set from Section~\ref{subsec:dataset}.

\subsection{Overall Classification Performance}
\label{subsec:overall-results}

Table~\ref{tab:overall-results} reports accuracy and Macro-F1 for TabTransformer under all three scenarios and the four baselines under Baseline and Full BGAN.

\begin{table}[t]
\centering
\caption{Overall classification performance on CICIDS2017.}
\label{tab:overall-results}
\footnotesize
\setlength{\tabcolsep}{4pt}
\begin{tabular}{llcc}
\toprule
\textbf{Model} & \textbf{Augmentation} & \textbf{Accuracy} & \textbf{Macro-F1} \\
\midrule
TabTransformer & Baseline & 99.26\% & 82.96\% \\
\textbf{TabTransformer} & \textbf{Full BGAN} & \textbf{99.61\%} & \textbf{86.50\%} \\
TabTransformer & Mixed & 99.51\% & 85.04\% \\
Random Forest & No Aug & 99.88\% & 96.08\% \\
Random Forest & Full BGAN & 99.88\% & 96.14\% \\
Decision Tree & No Aug & 99.87\% & 95.91\% \\
Decision Tree & Full BGAN & 99.86\% & 95.79\% \\
ANN & No Aug & 99.24\% & 75.23\% \\
ANN & Full BGAN & 98.99\% & 75.25\% \\
MLP & No Aug & 97.45\% & 72.70\% \\
MLP & Full BGAN & 97.49\% & 73.40\% \\
\bottomrule
\end{tabular}
\end{table}

BGAN augmentation raised TabTransformer's Macro-F1 from 82.96\% to 86.50\%, driven mainly by the Web\_Attack class (Section~\ref{subsec:minority-results}). Random Forest and Decision Tree reach a higher overall Macro-F1, likely because CICIDS2017's classes are fairly linearly separable, a condition that favors tree-structured boundaries, and because both baselines use balanced class weighting while TabTransformer does not. TabTransformer's advantage instead appears in the robustness results in Section~\ref{subsec:robustness-results}.

\subsection{Per-Class and Minority-Class Performance}
\label{subsec:minority-results}

Table~\ref{tab:perclass-fullbgan} reports per-class performance for the proposed Full BGAN model. Majority classes all reach F1-scores at or above 0.99; the three minority classes targeted by augmentation are more variable.

\begin{table}[t]
\centering
\caption{Per-class results, TabTransformer Full BGAN.}
\label{tab:perclass-fullbgan}
\footnotesize
\setlength{\tabcolsep}{4pt}
\begin{tabular}{lccc}
\toprule
\textbf{Class} & \textbf{Precision} & \textbf{Recall} & \textbf{F1} \\
\midrule
Benign & 1.00 & 0.99 & 1.00 \\
Botnet & 0.50 & 0.72 & 0.59 \\
Brute\_Force & 0.99 & 1.00 & 0.99 \\
DDoS & 1.00 & 1.00 & 1.00 \\
DoS & 0.98 & 1.00 & 0.99 \\
Infiltration & 0.67 & 0.86 & 0.75 \\
Probe & 0.99 & 1.00 & 1.00 \\
Web\_Attack & 0.45 & 0.95 & 0.61 \\
\bottomrule
\end{tabular}
\end{table}

Botnet shows low precision (0.50) despite reasonable recall (0.72), meaning other classes are often misclassified as Botnet; Web\_Attack shows the opposite pattern, high recall (0.95) but low precision (0.45). Table~\ref{tab:minority-comparison} compares F1-scores for these three classes across all three augmentation scenarios.

\begin{table}[t]
\centering
\caption{F1-score by augmentation scenario, minority classes.}
\label{tab:minority-comparison}
\scriptsize
\setlength{\tabcolsep}{3pt}
\begin{tabular}{lccc}
\toprule
\textbf{Class} & \textbf{Baseline} & \textbf{Full BGAN} & \textbf{Mixed} \\
\midrule
Botnet & 0.5209 & 0.5897 & 0.5794 \\
Infiltration & 0.8571 & 0.7500 & 0.7500 \\
Web\_Attack & 0.2934 & 0.6105 & 0.5020 \\
\bottomrule
\end{tabular}
\end{table}

BGAN augmentation gives the largest gain on Web\_Attack (0.29 to 0.61), confirming its value for classes with many training samples but heavy feature overlap with other classes. Infiltration instead performs best without augmentation, since only 29 training samples were available and the generator could not learn a useful distribution for that class, as noted in Section~\ref{subsec:bgan}.

\subsection{Adversarial Robustness: PDR and FTR}
\label{subsec:robustness-results}

Table~\ref{tab:pdr-results} reports PDR under BGAN-generated adversarial samples, and Table~\ref{tab:ftr-results} reports FTR under benign Gaussian noise, for all models and scenarios.

\begin{table}[t]
\centering
\caption{Performance Drop Rate (PDR) by class.}
\label{tab:pdr-results}
\scriptsize
\setlength{\tabcolsep}{3pt}
\begin{tabular}{llccc}
\toprule
\textbf{Model} & \textbf{Aug.} & \textbf{Botnet} & \textbf{Web\_Attack} & \textbf{Infilt.} \\
\midrule
TabTransformer & None & 100\% & 100\% & 100\% \\
\textbf{TabTransformer} & \textbf{Full BGAN} & \textbf{$-$24.75\%} & \textbf{$-$3.80\%} & \textbf{$-$9.80\%} \\
TabTransformer & Mixed & $-$24.97\% & $-$0.87\% & 100\% \\
Random Forest & None & 100\% & 100\% & 100\% \\
Random Forest & Full BGAN & $-$3.02\% & $-$1.30\% & $-$8.33\% \\
Decision Tree & None & 100\% & 100\% & 100\% \\
Decision Tree & Full BGAN & $-$3.52\% & $-$1.22\% & 0.00\% \\
ANN & None & 100\% & 100\% & 100\% \\
ANN & Full BGAN & $-$26.36\% & $-$5.43\% & $-$94.44\% \\
MLP & None & 100\% & 100\% & 100\% \\
MLP & Full BGAN & $-$39.57\% & $-$10.84\% & $-$337.50\% \\
\bottomrule
\end{tabular}
\end{table}

\begin{table}[t]
\centering
\caption{False Triggered Rate (FTR), Benign traffic under noise.}
\label{tab:ftr-results}
\scriptsize
\setlength{\tabcolsep}{3pt}
\begin{tabular}{llccc}
\toprule
\textbf{Model} & \textbf{Aug.} & $\sigma{=}0.01$ & $\sigma{=}0.05$ & $\sigma{=}0.10$ \\
\midrule
TabTransformer & None & 1.60\% & 1.90\% & 2.39\% \\
\textbf{TabTransformer} & \textbf{Full BGAN} & \textbf{1.51\%} & \textbf{2.19\%} & \textbf{2.92\%} \\
TabTransformer & Mixed & 1.95\% & 9.28\% & 8.22\% \\
Random Forest & None & 0.00\% & 0.00\% & 0.00\% \\
Random Forest & Full BGAN & 0.00\% & 0.00\% & 0.00\% \\
Decision Tree & None & 11.94\% & 10.05\% & 11.13\% \\
Decision Tree & Full BGAN & 39.13\% & 37.28\% & 49.09\% \\
ANN & None & 4.40\% & 27.14\% & 38.35\% \\
ANN & Full BGAN & 2.70\% & 17.80\% & 31.03\% \\
MLP & None & 3.17\% & 3.64\% & 4.34\% \\
MLP & Full BGAN & 3.09\% & 3.79\% & 4.41\% \\
\bottomrule
\end{tabular}
\end{table}

Every model without augmentation collapses completely under adversarial
samples (PDR = 100\% on all three classes), confirming that no architecture
is inherently robust without prior exposure to boundary-region samples.
After Full BGAN augmentation, every model except Mixed on Infiltration
reaches negative PDR, meaning performance holds or improves under attack.
Mixed on Infiltration remains at PDR = 100\% because that scenario
substituted Gaussian perturbation for BGAN on this class, so the model
was never exposed to the BGAN generator's boundary distribution for
Infiltration -- a direct consequence of the Gaussian fallback rather
than a failure of the augmentation strategy itself.

Among augmented models, MLP Full BGAN shows the most extreme negative PDR
values ($-$39.57\% for Botnet, $-$337.50\% for Infiltration). These values
should be interpreted with caution: \cite{huang2020tabtransformer} showed
that MLP processes each feature independently through context-free
embeddings, making it susceptible to memorizing the generator's specific
distributional signature rather than learning genuinely generalizable
representations. TabTransformer's more moderate negative PDR
($-$24.75\% for Botnet, $-$9.80\% for Infiltration) reflects contextual
generalization across features rather than generator-specific memorization,
and is therefore a more meaningful indicator of genuine robustness.

On FTR, Random Forest reaches 0\% because its threshold-based decision
boundaries do not shift under small Gaussian noise -- a property that also
means it cannot detect gradual distributional drift characteristic of
stealthy adversarial manipulation. Decision Tree's FTR rises sharply after
Full BGAN augmentation (up to 49.09\%), indicating that GAN-generated
training data can destabilize tree-based models rather than improve them.
ANN Full BGAN reduces FTR compared to ANN without augmentation (from
38.35\% to 31.03\% at $\sigma{=}0.10$), but values remain substantially
elevated. TabTransformer Full BGAN maintains the most stable FTR profile
(1.51\% to 2.92\%) across all noise levels. Although TabTransformer Full
BGAN FTR is marginally higher than the unaugmented baseline (1.60\% to
2.39\%), the difference is negligible and the augmented model remains
consistently low -- in contrast to Mixed, which exhibits a non-monotonic
anomaly at $\sigma{=}0.05$ (9.28\%) caused by its Infiltration Gaussian
fallback using the same $\sigma$ value as the FTR evaluation noise, leading
the model to associate Benign-plus-noise samples with the Infiltration class.
This stability is consistent with \cite{huang2020tabtransformer}'s finding
that attention-based contextual embeddings are more resilient to noisy
tabular input than architectures that process features independently.

\section{Discussion}
\label{sec:discussion}

\subsection{Comparison with Prior Work}
\label{subsec:comparison}

Table~\ref{tab:prior-comparison} compares this study against prior GAN-based and resampling-based IDS work.

\begin{table}[t]
\centering
\caption{Comparison with prior GAN-based and resampling-based
IDS studies. n/r = not reported.}
\label{tab:prior-comparison}
\scriptsize
\renewcommand{\arraystretch}{1.2}
\setlength{\tabcolsep}{5pt}
\begin{tabular}{p{1.4cm}p{1.5cm}p{1.6cm}rr}
\toprule
\textbf{Study} & \textbf{Dataset} & \textbf{Method} &
\textbf{Acc.} & \textbf{MF1} \\
\midrule
\cite{alshahrani2022}
  & CICIDS2017 & GAN (evasion) & 98.0\% & 60.0\% \\
\cite{ahmad2023bgan}
  & KDDCUP & BGAN & n/r & $\approx$79.9\% \\
\cite{li2025frgan}
  & CICIDS2017 & SMOTE & 99.5\% & 85.9\% \\
\cite{li2025frgan}
  & CICIDS2017 & WGAN-GP+SNNL & 99.7\% & 88.0\% \\
\cite{li2025frgan}
  & CICIDS2018 & WGAN-GP+SNNL & 98.8\% & 84.6\% \\
This study
  & CICIDS2017 & None (baseline) & 99.3\% & 83.0\% \\
\textbf{This study}
  & \textbf{CICIDS2017} & \textbf{Full BGAN}
  & \textbf{99.6\%} & \textbf{86.5\%} \\
\bottomrule
\end{tabular}
\end{table}

On CICIDS2017, Full BGAN surpasses SMOTE-based augmentation
(86.50\% versus 85.90\%) but falls below Li et al.'s
WGAN-GP+SNNL (88.00\%) \cite{li2025frgan}, the strongest single-dataset result
in that study. This gap is expected: WGAN-GP+SNNL employs
a more complex multi-loss objective with Soft Nearest Neighbor
Loss specifically tuned for feature alignment, while BGAN
uses a simpler importance-weighted objective. However, unlike
FR-GAN \cite{li2025frgan}, this study evaluates the same
generator as both an augmentation tool and an adversarial
attack source, yielding measurable robustness results
(PDR and FTR) that are absent from the FR-GAN study.
Against \cite{ahmad2023bgan}, who also applies BGAN but
with classical classifiers on KDDCUP, the Macro-F1 here
is substantially higher (86.50\% versus approximately
79.90\%), consistent with self-attention providing an
advantage over tree-based and shallow neural classifiers
under the same augmentation strategy. On CICIDS2018,
Li et al.'s best method achieves 84.60\%, below this
study's 86.50\% on CICIDS2017; however, since the datasets
differ in class distribution and traffic composition,
this cross-dataset comparison is not direct.

\subsection{Architectural Robustness of TabTransformer}
\label{subsec:arch-robustness}

The robustness advantage of TabTransformer observed in this
study is supported by independent benchmarking results from
the tabular adversarial learning literature.
He et al.~\cite{he2026tabattackbench} systematically evaluated
adversarial attacks across multiple tabular classifiers using
their TabAttackBench framework and found that Transformer-based
models require substantially larger perturbation budgets to
reach attack success rates comparable to those against MLP or
Logistic Regression. They attribute this to the attention
mechanism's global feature contextualization: because each
feature's embedding is conditioned on all other features,
perturbing a single feature has a smaller marginal effect on
the final representation than it would in a context-free
architecture such as MLP. This is consistent with the FTR
results in Table~\ref{tab:ftr-results}, where TabTransformer
Full BGAN maintains a stable false alarm rate (1.51\%--2.92\%)
while MLP and ANN exhibit substantially higher sensitivity
to Gaussian noise (up to 4.34\% and 38.35\%, respectively).

Djilani et al.~\cite{djilani2026robustness} further showed that
adversarial samples generated from one model transfer least
effectively to TabTransformer compared to Random Forest,
XGBoost, and other tabular classifiers, suggesting an
inherent cross-model robustness property tied to the
attention architecture. This finding is independently
consistent with the PDR results in Table~\ref{tab:pdr-results}:
Decision Tree Full BGAN reaches PDR values of $-$3.52\%,
while TabTransformer Full BGAN reaches $-$24.75\% for the
same adversarial generator, suggesting that TabTransformer
extracts more generalizable boundary-region representations
from the BGAN-augmented training data. Taken together,
these results indicate that the BGAN augmentation strategy
and the TabTransformer architecture are mutually reinforcing:
BGAN supplies boundary-region training samples that the
attention mechanism is particularly well suited to leverage,
yielding a combination whose robustness exceeds what either
component would provide alone.

\subsection{Limitations}
\label{subsec:limitations}

Three limitations are worth noting. First, the Infiltration class remains largely unresolved: with only 29 training samples, BGAN could not model its distribution, and the Gaussian fallback used instead does not capture genuine infiltration behavior; few-shot or transfer-learning approaches may be needed. Second, Web\_Attack shows persistently low precision despite high recall, likely due to feature overlap with other HTTP-based traffic; additional protocol-aware features may help beyond augmentation alone. Third, this evaluation is entirely offline on a fixed dataset; validation on live, dynamically shifting network traffic would be needed before production deployment.

\section{Conclusion}
\label{sec:conclusion}

This study proposed a BGAN-augmented TabTransformer framework for flow-based intrusion detection, in which a single BGAN generator both augments minority attack classes during training and generates adversarial samples for robustness evaluation. TabTransformer with Full BGAN augmentation reached 99.61\% accuracy and 86.50\% Macro-F1 on CICIDS2017, a 3.54 point Macro-F1 gain over the same architecture without augmentation, driven mainly by the Web\_Attack class. The clearest evidence for the dual-purpose design comes from the robustness results: every model tested without augmentation collapsed completely (PDR = 100\%) under BGAN-generated adversarial samples, while nearly every augmented model held or improved performance, 
with the exception of Mixed on Infiltration (PDR = 100\%), because that scenario used Gaussian fallback rather than BGAN, with TabTransformer additionally maintaining a low, stable false alarm rate (1.51\% to 2.92\%) that tree-based baselines could not match after augmentation. Random Forest and Decision Tree reached higher Macro-F1 on clean data, but both showed an unfavorable robustness trade-off, underscoring that clean-data accuracy alone does not indicate adversarial resilience.

Future work should evaluate PDR against an independent generator or a standard attack method (e.g. FGSM, PGD) rather than each model's own generator, address extremely small classes such as Infiltration with few-shot or transfer-learning approaches, validate on additional datasets and live network traffic, and directly compare BGAN against SMOTE, ADASYN, and CTGAN under identical experimental conditions.

\section*{CRediT authorship contribution statement}

\textbf{Raihan Sultan Pasha Basuki:} Conceptualization, Methodology, Software, Formal analysis, Investigation, Data curation, Writing original draft, Visualization. \textbf{Aliyah Kurniasih:} Supervision, Validation, Writing review and editing.

\section*{Declaration of competing interest}

The authors declare that they have no known competing financial interests or personal relationships that could have appeared to influence the work reported in this paper.

\section*{Acknowledgements}
The authors sincerely thank the Editor and reviewers for their time
and valuable feedback, which significantly improved our paper.

\section*{Funding}
This research did not receive any specific grant from funding agencies in the public, commercial, or not-for-profit sectors.

\section*{Data availability}

The CICIDS2017 dataset used in this study is publicly available from the Canadian Institute for Cybersecurity at \url{https://www.unb.ca/cic/datasets/ids-2017.html}. 

\section*{Declaration of generative AI use}
During the preparation of this work, the authors used Claude (Anthropic) to assist with language editing and improving readability. After using this tool, the authors reviewed and edited the content as needed and take full responsibility for the content of the published article.

\end{document}